# Target-Aware Implementation of Real Expressions

Brett Saiki
University of Washington
Seattle, WA, USA
bsaiki@cs.uw.edu

Jackson Brough
University of Utah
Salt Lake City, UT, USA
jackson.brough@utah.edu

Jonas Regehr
University of Utah
Salt Lake City, UT, USA
u0760638@utah.edu

Jesús Ponce
University of Utah
Salt Lake City, UT, USA
u1398911@utah.edu

Varun Pradeep
University of Washington
Seattle, WA, USA
varun10p@cs.washington.edu

Aditya Akhileshwaran
University of Washington
Seattle, WA, USA
aakhiles@cs.washington.edu

Zachary Tatlock
University of Washington
Seattle, WA, USA
ztatlock@cs.uw.edu

Pavel Panchekha
University of Utah
Salt Lake City, UT, USA
pavpan@cs.utah.edu

## Abstract

New low-precision accelerators, vector instruction sets, and library functions make maximizing accuracy and performance of numerical code increasingly challenging. Two lines of work—traditional compilers and numerical compilers—attack this problem from opposite directions. Traditional compiler backends optimize for specific target environments but are limited in their ability to balance performance and accuracy. Numerical compilers trade off accuracy and performance, or even improve both, but ignore the target environment. We join aspects of both to produce Chassis, a target-aware numerical compiler.

Chassis compiles mathematical expressions to operators from a target description, which lists the real expressions each operator approximates and estimates its cost and accuracy. Chassis then uses an iterative improvement loop to optimize for speed and accuracy. Specifically, a new instruction selection modulo equivalence algorithm efficiently searches for faster target-specific programs, while a new cost-opportunity heuristic supports iterative improvement. We demonstrate Chassis' capabilities on 9 different targets, including hardware ISAs, math libraries, and programming languages. Chassis finds better accuracy and performance trade-offs than both Clang (by 3.5×) or Herbie (by up to 2.0×) by leveraging low-precision accelerators, accuracy-optimized numerical helper functions, and library subcomponents.

## 1 Introduction

Today's programming environments increasingly provide hardware accelerators, special numeric software libraries, and numeric programming languages [37]. Programmers must use these new capabilities to maximize the performance of their programs, but their accuracy characteristics are esoteric, raising the risk of catastrophic numerical error. Writing numerical code requires ever more expertise on increasingly heterogeneous environments to preserve correctness and achieve performance.

We can reframe this challenge as a compilation problem: implementing numerical code from mathematical formulas is just translation of high-level descriptions to executable programs. Today, two different styles of compilation exist to address the challenge. Traditional compilers like Clang specialize programs to a wide diversity of targets. However, they do little with regard to accuracy, either preserving bit-identical results (which typically prevents accelerators from being used) or allowing unrestricted "fast-math" optimizations that disregard numerical accuracy entirely. On the other hand, numerical compilers such as Herbie [29, 31], Rosa [12], and others [5, 8, 30] utilize precision tuning [8, 12, 30], term rewriting [29], or both [31] to improve accuracy and performance. However, existing numerical compilers are target-agnostic and can't specialize to the intended target. To bridge this gap, we combine accuracy-aware term rewriting—not found in traditional compilers—with target-aware instruction selection which is missing from numerical compilers.

The result is Chassis, a target-aware, numerical compiler that optimizes for both accuracy and performance. Chassis compiles real-number expressions to a set of target-specific floating-point programs that optimize for performance at various accuracy bounds (a Pareto frontier). Compared to traditional compilers, Chassis exploits mathematical identities without causing the catastrophic accuracy losses for which "fast-math" optimizations are notorious [7]. Compared to numerical compilers, Chassis allows developers to select a specific target environment and leverages target-specific instructions in generated floating-point programs, thereby achieving better accuracy and performance.

To do so, Chassis uses a core *operator* abstraction, which relates target-specific operations to the real number expressions they approximate. This insight permits a novel algorithm for *instruction selection modulo equivalence* that uses equality saturation on mixed real-float expressions to select target-specific instructions, preserve mathematical equality, and maximize performance. To direct this heavyweight optimization pass, Chassis uses an iterative improvement loop guided by *accuracy- and cost-based analyses* that identify inaccurate or slow regions of code. As a result, Chassis can quickly identify problematic expressions, improve numerical error and performance, and make target-specific optimization decisions.

We demonstrate Chassis on a diverse set of 9 targets covering hardware ISAs, programming languages, and math libraries. We show that, for these targets, Chassis achieves better accuracy and performance than state-of-the-art traditional and numeric compilers. On our C 99 target, Chassis attains a speed-up at equivalent accuracy of 8.9× over Clang. On our Python and vdt library targets, Chassis achieves speed-ups of up to 2.0× and up to 1.9× over Herbie. Moreover, these improvements are due to Chassis' ability to adapt to target-specific characteristics.

This paper contributes the following:

- A target description language that relates floating-point instructions to the real expression that they approximate (Section 4).
- An algorithm for instruction selection modulo equivalence leveraging equality saturation (Section 5.1).
- Accuracy and cost heuristics to guide this heavyweight optimization toward the most significant cases of numerical error and bad performance (Section 5.2).

Section 2 provides additional motivation and an overview of Chassis. Section 3 follows with necessary background. Section 4 presents Chassis' target description language while Section 5 describes Chassis' target-specific optimization strategy. Section 6 evaluates Chassis against existing traditional and numerical compilers. Section 7 discusses Chassis' limitations and how future work may address them. Section 8 presents related work, and Section 9 concludes.

## 2 Overview

Consider writing an inverse hyperbolic cotangent function,

$$\coth^{-1}(x) = \frac{1}{2}\left[\log\left(1 + \frac{1}{x}\right) - \log\left(1 - \frac{1}{x}\right)\right],$$

for three different targets: the AVX extensions to the x86 ISA, the Julia language, and Sun Microsystems' fdlibm library. Each of these targets has its own complex concerns.

A numerical programmer targeting AVX should maximally leverage instructions like AVX's fused multiply-add/subtract operations ($xy + z$, $xy - z$, $-xy + z$, and $-xy - z$) since these instructions are more accurate and faster than separate multiply and add/subtract steps. Additionally, AVX lacks a negation instruction,[1] so they should fold negations into multiply-add/subtracts. Ideal AVX code may also leverage AVX's rcpps and rsqrtps instructions, which compute $1/x$ and $1/\sqrt{x}$ faster, but less accurately, than the division or square root operations. Actual division and square root operations are fairly slow (latencies of 11 and 12 cycles, compared to just 4 for fused multiply-add [20]), so they are best avoided unless maximum accuracy is needed. Transcendental functions like $\sin(x)$ and $e^x$ aren't available; AVX code must use polynomial approximations instead. Finally, since conditionals are implemented in AVX by masking, ideal AVX code should use "if-else" expressions only in cases where they are critical for accuracy.

---

[1] One can use integer operations to flip the sign bit, but this usually requires an extra register, so isn't ideal.

A numerical programmer targeting Julia, a widely used language for numerical analysis and scientific computing, has very different concerns. Julia offers a normal scalar execution model, so they can use conditionals freely; additionally, Julia code has substantial overheads, so the added cost of division and square root is nearly unobservable. Unlike AVX, Julia supports not only transcendental functions but also an extended math library with helper functions for $\sin(\pi x)$, $\log(1 + x)$, $\sqrt{x^2 + y^2}$, and others. Maximally using these helper functions not only adds clarity but also accuracy, since these helpers perform operations like multiplication by $\pi$ in higher internal precision.

Finally, a numerical programmer using fdlibm, Sun's implementation of the standard C math library, can leverage their knowledge of the library's internals for greater speed. For example, fdlibm implements the $\log(x)$ function by first normalizing the exponent of $x$ through range reduction, then transforming $x$ into a related value, $s = (x - 1)/(x + 1)$, for which it computes $\log(1 + s) - \log(1 - s)$ accurately. If a computation can be rewritten to use $\log(1 + s) - \log(1 - s)$ (which we call log1pmd) instead of log, we may achieve a speedup. Today, calling library-internal subroutines is the domain of true numerical experts, but automating it can provide substantial speedups.

Because these targets are so different, a moderate-accuracy inverse cotangent implementation looks different on each. On AVX, it would use a polynomial approximation evaluated at $1/z$, with $1/z$ computed with the fast reciprocal instruction; in Julia, it would use the log1p helper functions; and in fdlibm, it would directly call the log1pmd subroutine. Note that these different implementations are not just ports of some target-independent implementation: the best possible speed and accuracy can be achieved only by understanding the underlying target's numerical capabilities and performance characteristics.

Chassis' main goal is to account for this *wide diversity* of target characteristics, including differing sets of instructions, operator accuracies, relative latencies, conditional execution style, unusual operators, and more. Chassis therefore provides a *target description language* (Section 4) that captures these features and uses its target descriptions to synthesize and apply *target-specific optimizations* (Section 5).

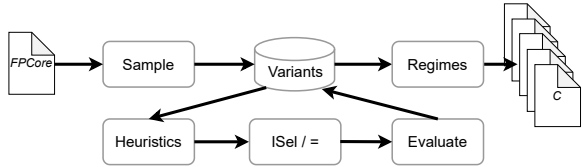

**Figure 1.** Chassis' system architecture. Chassis takes an input program and a target description (not shown) and samples input points for measuring accuracy. Using heuristics, it iteratively applies its instruction selection algorithm to explore many possible implementations, keeping only those with high accuracy or performance. Finally, Chassis extracts multiple Pareto-optimal programs.

Chassis system is illustrated by Figure 1. The frontend parses an input expression (in the standard FPCore format [11]) and a target description. Using the expression, it samples training and test inputs that Chassis uses to evaluate accuracy. It then enters an iterative loop, described below, for a fixed number of iterations, which generates a set of candidate programs with varying Pareto-optimal speeds and accuracies. Finally, Chassis combines candidates using conditionals, and outputs the programs in either a target-specific format or in the default FPCore format. This overall architecture, as well as the sampling and regime steps, are shared with prior work [29, 31] and do not require significant changes to support target-specificity.

Chassis' contributions are found in its iterative loop. Each iteration has three phases. First, Chassis uses its accuracy- and cost-aware heuristics to identify a small set of subexpressions that should be rewritten for improved speed or accuracy. Then, a heavier-weight rewriting pass[2] using our instruction selection algorithm generates new variations of that subexpression, which are substituted to produce new program variants. Finally, Chassis evaluates the cost and accuracy of each variant using the training inputs and the target's accuracy model and retains the Pareto-optimal choices, that is, the most accurate programs for any given accuracy bound. To start the next iteration, Chassis chooses one or more of these retained variants and repeats the process with their subexpressions.

## 3 Background

We provide a brief, necessary background for understanding the details of Chassis' implementation. First, we compare Chassis to prior work (Section 3.1). Then, we cover the e-graph data structure (Section 3.2) and equality saturation (Section 3.3), both of which Chassis uses during optimization.

---
[2]As in prior work [29], series expansion also generates variants.

### 3.1 Herbie

Chassis' system architecture borrows heavily from the Herbie numerical compiler [29]. Herbie compiles mathematical expressions to high-accuracy floating-point programs, with later extensions [31] producing multiple outputs. It parses an input expression in FPCore, samples training and test inputs, and iteratively builds a set of candidate programs. Each iteration has three steps. It uses *local error* [29] to determine the subexpressions introducing the most error. Then, it applies various rewriting strategies (mathematical identities, series expansion, and precision tuning) to generate many candidates. Finally, it keeps those with the best accuracy for a given cost budget on at least one input. After a fixed number of iterations, Herbie exits the loop and executes its regime inference algorithm [29] which selects a subset of programs that perform well on different parts of the domain and fuses these programs together with branch conditions.

To evaluate the accuracy of a program on an input, Herbie compares the floating-point result to the "correctly-rounded" result, the nearest floating-point number when computed using real-number semantics. Specifically, correct rounding comes from the Rival interval arithmetic library [19], which supports algebraic, exponential, logarithmic, and trigonometric operators, as well as hyperbolic, modulus, and rounding operators. Herbie produces programs that use exactly the operators from Rival (possibly at different precisions).

Unlike Chassis, Herbie does not consider target characteristics. Where it considers cost, Herbie uses a mix of two different, simplistic cost models: one which simply assigns a cost of 1 for each operation, and another where arithmetic has cost 1 and all other function calls have cost 100. One can think of this naive cost model as approximating a wide range of hardware and software targets where Herbie only requires *relative performance* to rank candidates. Chassis uses a similar system architecture, but with different, target-aware localization and rewrite phases which allow it to produce substantially faster code.

### 3.2 E-graphs

An e-graph is a data structure that efficiently represents an equivalence relation over a set of terms. Its advantage is that a small e-graph can often represent a large, exponential, or even infinite set of equivalent terms. Internally, an e-graph is a set of *e-classes* (equivalence classes) where each e-class is a set of equivalent *e-nodes*; an e-node $f(c_1, \ldots, c_n)$ is a function symbol $f$ with a list of children e-classes $c_i$. Two terms represented by the same e-class are considered equivalent. If the e-graph contains a cycle, it represents an infinite set of terms. Additionally, e-graphs perform deduplication operations to remain compact.

Figure 2 shows a small e-graph containing the equivalent terms $x + x$, $2x$, and $x \ll 1$. Note particularly that the "arguments" in an e-node are e-*classes*, not e-nodes. Thanks to this, a single e-node can represent a large set of terms, which gives the e-graph representation its power. Formally, an e-graph represents a term $t$ if any e-class represents $t$; an e-class represents $t$ if any of its e-nodes represent it; an e-node $f(c_1, \ldots, c_n)$ represents a term $f(t_1, \ldots, t_n)$ if each $c_i$ represents $t_i$.

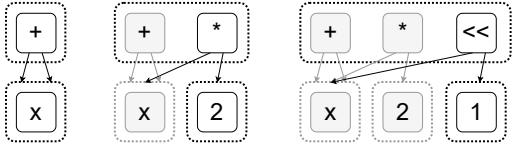

**Figure 2.** Initially, an e-graph (left) contains the term $x + x$. The rewrites $a + a \rightsquigarrow 2 * a$ (middle) and $2 * a \rightsquigarrow a \ll 1$ (right) are applied, introducing new terms. The dotted boxes represent e-classes while the solid boxes represent e-nodes.

### 3.3 Equality Saturation

Equality saturation [35] uses e-graphs to optimize programs. It constructs an e-graph representing an initial program $p$ and incrementally applies rewrite rules $\ell \rightsquigarrow r$ which expand the e-graph. Users of equality saturation, which include Chassis, can then *extract* some specific program $p'$ from the e-class of $p$ in the e-graph. If each rewrite rule $\ell \rightsquigarrow r$ preserves some equivalence relation, then $p$ and $p'$ are equivalent as well. A number of different extraction algorithms exist, including greedy algorithms [39] or integer linear programming [24, 35, 38].

Importantly, equality saturation applies rewrite rules *non-destructively*. For $\ell \rightsquigarrow r$, the algorithm uses *e-matching* [17], to find terms matching $\ell$, say $\sigma(\ell)$ in e-class $c_\ell$. The substitution $\sigma$ creates a new term $\sigma(r)$, equivalent to $\sigma(\ell)$, and its e-class is merged with $c_\ell$. If $\sigma(r)$ is not already represented, a fresh e-class is made for it before merging. Note that $\sigma(\ell)$ is not removed. This non-destructive property allows equality saturation to explore many compositions of rewrite rules in parallel, largely circumventing the phase-ordering problem [35].

Equality saturation can be useful for traversing the space of mathematically-equivalent formulas. For example, Herbie uses a database of 325 hand-written rewrite rules for equality saturation. Herbie transforms $\sqrt{x+1} - \sqrt{x}$ using the rule $a - b \rightsquigarrow (a^2 - b^2)/(a+b)$, as well as rules including $a + b \rightsquigarrow b + a$, $(a + b) - c \rightsquigarrow a + (b - c)$, and $\sqrt{a}^2 \rightsquigarrow a$ to arrive at $1/(\sqrt{x+1} + \sqrt{x})$. This new expression is equivalent to the original over the real numbers but is much more accurate when evaluated in floating-point.

## 4 Target Description Language

Chassis' core abstraction, the *operator*, allows it to handle compilation targets with different numeric capabilities, and perform a variety of numeric optimizations.

### 4.1 Desugaring and Lowering

Operators are the atomic instructions of Chassis' internal IR; they take floating-point inputs (of a given type) and produce floating-point outputs. Outputs are pure and total; error cases are modeled as returning NaN. Crucially, each operator has a *desugaring*, or denotation: a real-number expression that defines its semantics. Chassis optimizations ensure that the desugaring of the input program, but not necessarily its floating-point semantics, is preserved.[3]

This desugar-preservation semantics subsumes existing numerical optimizations. Operations may have simple desugarings such as $\exp_{f64}(a) \rightsquigarrow e^a$, or complex desugarings like $\text{log1pmd}_{f64}(a) \rightsquigarrow \log(1+a) - \log(1-a)$. In either case, Chassis can rewrite floating-point programs by applying mathematical identities based on their desugarings as in Herbie, Pherbie, and Salsa [10, 29, 31]. A desugaring is not necessarily unique to a particular program. For example, the AVX programs $1/_{f64}x$ and $\text{rcpps}(x)$ both map to $1/x$; Chassis can then preserve the desugaring by selecting either option, as needed for speed or accuracy. Two operators may be even more closely related: vdt's $\sin(x)$ and $\text{fast\_sin}(x)$ only have different accuracy and performance. In such cases, Chassis is allowed to perform implementation selection like in OpTuner [5]. An operator can even have a trivial desugaring: an operation casting $a$ from double to single precision desugars to just $a$. Desugaring preservation then means precision tuning à la FPTuner [8]. In short, Chassis' operator abstraction, with its notion of desugaring preservation, provides a simple compiler-internal IR that allows the most important speed and accuracy optimizations in numerical code.

Note that while each floating-point program has a deterministic desugaring, a mathematical expression can have any number of *lowerings* back to floating-point programs. For example, we may implement $\log(1 + \sqrt{1 + x^2})$ in floating-point code directly as-is, or Chassis could use library functions like $\text{log1p}_{f64}(a) \rightsquigarrow \log(1 + a)$ or $\text{hypot}_{f64}(a, b) \rightsquigarrow \sqrt{a^2 + b^2}$ to produce $\text{log1p}_{f64}(\text{hypot}_{f64}(1, x))$. Some lowerings happen by way of algebraic rewrites: the real expression $x/y$ is mathematically equivalent to $x \cdot (1/y)$, so it can be lowered to $\text{mulps}(x, \text{rcpps}(y))$. This ambiguous lowering drives Chassis' optimization algorithm: Chassis both rewrites the input program using mathematical identities and substitutes in operators available on a given target to find fast and accurate output programs.

### 4.2 Target Descriptions

The operator abstraction guides Chassis' target description language: a Chassis target description is a list of available operators, where each operator has a name, a type signature, and a desugaring, all expressed in a simple S-expression DSL. Figure 3 shows a fragment of Chassis' AVX target description. Since Chassis focuses on the interplay between speed and accuracy of floating-point operations, targets only describe floating-point operations. Traditional compiler concerns such as integer and boolean operations, control flow, memory, and calling conventions are left to some target-specific compiler invoked on the output of Chassis.

---

[3]Chassis also uses polynomial approximations; MegaLibm-style [6] approx terms could be added to Chassis to maintain this same desugaring-preservation semantics.

```
1   (define-operator (rcp.f32 [x binary32]) binary32
2    #:approx (/ 1 x)
3    #:link (lib "libavx" rcpps)
4    #:cost 4.0)
5
6   (define-operator (/.f32 [x binary32] [y binary32]) binary32
7    #:approx (/ x y)
8    #:cost 10.0)
9
10  (define-target avx
11   #:if-cost (max 5)
12   #:literals ([binary32 1])
13   #:operators [rcp.f32 /.f32 . . . ])
```

**Figure 3.** Part of Chassis' AVX target description. Operators require a desugaring and linking for evaluation, as well as a scalar cost. Target descriptions list the available operators and provide other cost model information.

Besides the set of available operators, Chassis must also estimate the accuracy and speed of generated programs. For accuracy, Chassis executes each operation on training inputs; to support this, target descriptions specify a symbol in a shared library, which Chassis will dynamically link to for evaluation (line 3 in Figure 3). This implementation can run the corresponding instruction on the host system (as with the AVX target) or simulate its behavior (for cross-compilation). To estimate speed, Chassis target descriptions provide a scalar cost for each operation, where the speed of a program is assumed to be inversely related to the sum of costs of each operator, plus customizable costs for literals and variable references. Targets can also specify how Chassis evaluates the cost of a conditional: "scalar"-style targets evaluate the cost of the predicate plus the cost of the more expensive branch, while "vector"-style targets evaluate the cost of the predicate plus the cost of both branches. Our experiments show that Chassis' cost models, despite their simplicity, suffice for the fairly coarse distinctions needed to rank Chassis' generated candidates. Finally, a target description can optionally describe the output format (e.g., C) that Chassis output programs in, and a compiler driver to compile them.

Importantly, we find that Chassis can achieve good results even with *highly approximate* cost and accuracy information.

Thus, Chassis can auto-generate accuracy and speed information, which makes developing new target descriptions straightforward. If the user does not provide linking information, Chassis will *synthesize* a correctly-rounded—meaning, maximally accurate—implementation using Rival (/.f32 in Figure 3). These synthesized implementations are typically good enough for helper functions like hypot or log1pmd, though linking information is necessary for approximate operators like rcpps. If the user does not provide cost information, Chassis provides an auto-tuner that estimates the cost of each operator by compiling and measuring the runtime of short programs that call that operator in a hot loop.[4] Naturally, these auto-tuned costs are not very accurate, but seem to suffice in practice (see Section 7 for further discussion). More concisely, while providing linking and cost information is necessary to get the best results from Chassis, the default correctly-rounded and auto-tuned data makes it possible to develop new targets quickly.

To further ease development of new targets, targets can import, combine, or modify other targets. For example, if a "core C" target describes the basic arithmetic operations (+, −, ∗, /), a "libm" target may import the core C target and add more operators for various math library functions. Importantly, the target for the fdlibm library can import the same core C target and add the same math functions, but with different costs and linking to different implementations. This ease of target development not only allows Chassis users to keep up with increasingly heterogeneous programming environments, but also allows them to inform Chassis about libraries and helper functions which Chassis can then leverage.

## 5 Optimization and Lowering

Chassis *iteratively* uses e-graphs in its novel *instruction selection algorithm* to rewrite floating-point programs while preserving the desugaring of the initial program thereby generating fast and accurate implementations.

### 5.1 Instruction Selection Modulo Equivalence

The core insight behind Chassis' optimization pass is that desugaring preservation is an equivalence relation: two floating-point programs are equivalent in Chassis when their desugarings are found to be equivalent either syntactically or by mathematical identities. This allows Chassis to use equality saturation via the egg library [39] for its core optimization pass, *instruction selection modulo equivalence*. At its core, instruction selection modulo equivalence is traditional tree-pattern matching, but performed over an e-graph describing an equivalence relation.

**Mixed real/float expressions.** Existing numerical compilers restrict equality saturation to floating-point programs at a fixed precision, using rewrite rules like $(a +_{\text{f32}} b) +_{\text{f32}} c \leadsto a +_{\text{f32}} (b +_{\text{f32}} c)$. Precision changes are performed separately [31]. This approach is insufficient for Chassis. First, each target description would require its own set of rewrite rules over *floating-point terms* to accommodate target-specific operations, making target definitions hard to write. Second, some target-specific operators, such as those in Intel's AMX extensions, require specific argument precisions that are difficult to achieve if rewrites and precision changes happen separately. Finally, if a target lacks certain operations, critical transformations may be difficult to express, or even discover, through composition [27, 28].

---

[4]This requires the user to specify an output format and compiler for the target; if they do not, Chassis defaults to assigning a cost of 1 per operator.

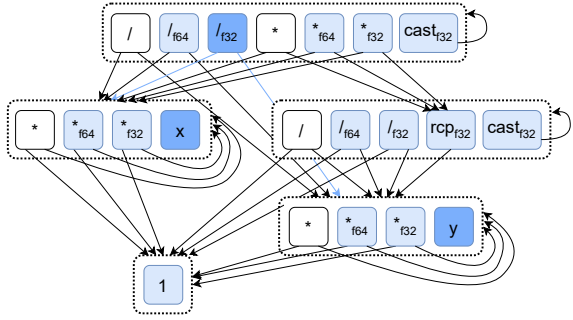

**Figure 4.** An e-graph after applying instruction selection modulo equivalence to $x/_{\text{f32}} y$ (dark blue). The e-graph contains a *mix* of real-number e-nodes (white) and floating-point e-nodes (blue), where each e-class represents equivalence under real semantics. Extraction selects a well-typed floating-point program that is minimal according to the target description's cost model; extraction considers all blue nodes.

Chassis instead performs equality saturation over *mixed real-float* expressions. Specifically, Chassis constructs an equality saturation problem combining mathematical equivalences like $a/b \leadsto a \times (1/b)$ and desugaring equivalences like $a/_{\text{f32}} b \leadsto a/b$ and $1/a \leadsto \text{rcp}_{\text{f32}}\, a$ drawn *directly* from the target description. This naturally introduces mixed real-float expressions like $x \times_{\text{f32}} (1/y)$, shown in Figure 4. In this representation, mathematical equivalences are defined once, and do not have to be specialized to each target. Precision tuning can also be performed in the same equality saturation pass as rewrites, making mixed-precision accelerators easy to generate. Moreover, Chassis can use mathematical identities even when they contain real-number operators that have no direct analog to target-specific operations.

**Typed extraction.** However, mixed real-float expressions introduce a new problem: the e-graph now includes not only valid floating-point programs but also ill-typed programs (like $x \times_{\text{f32}} (1/_{\text{f64}} z)$), mixed real/float programs (like $x \times_{\text{f32}} (1/z)$), and real programs with no valid desugaring (on some targets, $x \times \text{pow}(z, -1)$). Chassis must avoid extracting any such programs from the e-graph.

To solve this problem, Chassis implements a novel *typed extraction* algorithm. Like existing e-graph extraction algorithms, typed extraction tracks and updates the lowest-cost expression in each e-class until fixed-point is reached. However, unlike existing extraction algorithms, typed extraction restricts the results to well-typed expressions and tracks one lowest-cost expression for every possible floating-point type. For each e-node, typed extraction looks up the cheapest expression of the appropriate type from each argument e-class, and adds their costs to the operator cost to get a cost for the e-node[5]. If any argument e-class does not have an expression of the appropriate type, the e-node is ignored. Once fix-point is reached, Chassis can easily output the lowest-cost program of the same floating-point type in the same e-class as the initial program.

To illustrate how the algorithm works, consider an e-class containing e-nodes $c_1/c_2$, $c_3/_{\text{f64}} c_4$, $c_5/_{\text{f32}} c_6$, and $\text{rcp}_{\text{f32}}(c_7)$. Typed extraction ignores $c_1/c_2$ since it uses a real-number operator, and then groups the remaining e-nodes by output type. The double-precision e-node $c_3/_{\text{f64}} c_4$ is considered separately from the single-precision e-nodes $c_5/_{\text{f32}} c_6$ and $\text{rcp}_{\text{f32}}(c_7)$, and typed extraction will extract (at most) one lowest-cost expression from each. For $c_3/_{\text{f64}} c_4$, typed extraction finds the simplest double-precision expressions from $c_3$ and $c_4$, $e'_3$ and $e'_4$, and uses the term $e'_3/_{\text{f64}} e'_4$ as the lowest-cost double-precision output. For single precision, the two e-nodes are extracted in the same fashion, and their costs compared to determine which has the lowest cost.

### 5.2 Iterative Improvement

It is tempting to have Chassis simply perform instruction selection modulo equivalence and output the results. However, while this algorithm excels at lowering cost, it does not consider accuracy, and tends to introduce unacceptable numerical error. Drawing from the literature on numerical compilers, Chassis instead uses an iterative, heuristic-guided search, first generating a large number of programs, then evaluating them for *both* accuracy and cost to keep the Pareto-optimal subset, and finally repeating to find programs with even better accuracy and performance.

**Multi-extraction.** Extracting a single output program from the e-graph would over-optimize for performance at the cost of accuracy. The natural fix is to extract many output programs from the e-graph and retain the ones with the best performance/accuracy trade-off. However, since the e-graph contains an exponentially-large (or even infinite) set of programs, extracting and ranking all programs in the e-graph is infeasible. Chassis addresses this problem in two ways. First, Chassis extracts every appropriately-typed e-node instead of just the lowest-cost one. Second, it does so only for those subexpressions in the original program that introduce the most error or cost. When these e-nodes refer to other e-classes, the lowest-cost appropriately-typed option is used. Together, these two tweaks[6] ensure Chassis considers a large number of candidate programs but only in the part of the program where speed or accuracy improvements are likely. This makes the process feasible: there is now at most one extracted expression per e-node in the e-graph[7], even when the e-graph represents an exponentially-large or infinite set of expressions[8].

---

[5]Note that the types and costs of each operator are available in the target description.

**Heuristic Guidance.** Chassis chooses subexpressions of interest using its *local error* and *cost opportunity* heuristics. The local error heuristic was introduced in Herbie [29] and developed further in Hi-FP-Tuner [22]: it identifies subexpressions with high error. Importantly, the error an operator introduces is measured in *isolation* from its children, meaning that operators aren't blamed for errors introduced by their arguments. Instead, Chassis' cost opportunity heuristic identifies subexpressions with high cost; however, this is not straightforward. Some expensive operations, such as transcendental functions, cannot be performed in any low-cost way. Cost opportunity analysis must predict where rewrites could be valuable.

Chassis' cost opportunity heuristic does this using a fast, lightweight equality saturation pass. Cost opportunity analysis rewrites all subexpressions of the input program using a limited set of "simplifying" rewrites (mathematical identities that do not grow the AST), then computes the lowest-cost output for each. Cost opportunity then compares to original and simplified expressions for each subexpressions. The difference in their costs, minus that same difference for their arguments, is the cost opportunity of that node. Subtracting the cost difference for arguments ensures that a node is not "credited" for simplifications that must be performed on their arguments.[9] Figure 5 shows the cost opportunity algorithm. Since the set of rewrites is smaller, the analysis is much faster than the actual rewrite pass, despite applying to more subexpressions.

As an example, consider $1 +_{f32} (x /_{f32} y)$. It has two (non-trivial) subexpressions: the addition and the division. Suppose the simplest form for both subexpressions comes from $1 +_{f32} (x *_{f32} \text{rcp}_{f32}(y))$. The cost opportunity for the division, then, is the cost of $x /_{f32} y$ (about 11 cycles of latency), minus the cost of $x *_{f32} \text{rcp}_{f32}(y)$, a gap of about 4 cycles. The cost opportunity of the addition, however, is zero. Although the simplified form of that subexpression has a lower cost, the cost savings are from the children and are thus subtracted

---
[6]These changes are inspired by the MegaLibm compiler [4, 6].
[7]At most one, because many e-nodes refer to ill-typed expressions and are not extracted.
[8]Chassis limits e-graphs to 8 000 nodes, and typically extracts about 40 candidate programs for each subexpression of interest.
[9]If this consideration is ignored, the root of the program will typically have the largest cost opportunity, since it is credited for any optimization performed anywhere in the program.

```
1  def cost_delta(E, e):
2      return cost(e) - cost(extract(E, e))
3
4  def cost_opportunity(e):
5      E = run_egraph(e)
6      Δcost_e = cost_delta(E, e)
7      Δcost_children = sum(cost_delta(E, c) for c in e.children())
8      return Δcost_e - Δcost_children
```

**Figure 5.** Cost opportunity analysis. Simplifying rewrites are applied to every subexpression. The function cost applies the target-specific cost model to the given expression. For each subexpression, the cost opportunity is the difference in costs between its original and simplified form minus the difference between its subexpressions' cost.

out. As a result, Chassis correctly focuses its rewrites on the division operation. Cost opportunity is a useful heuristic for identifying expressions where rewriting could lead to much faster expressions. Chassis uses cost opportunity in conjunction with local error to select subexpressions to perform optimizations on.

## 6 Evaluation

We evaluate four research questions:

1. Can Chassis compile to a diverse set of targets? (Section 6.1)
2. Does Chassis produce faster code, for a given accuracy, than the traditional compiler Clang? (Section 6.2)
3. Does Chassis produce faster code, for a given accuracy, than the numerical compiler Herbie? (Section 6.3)
4. Does Chassis produce faster code because its optimizations are target-specific? (Section 6.4)

We performed all experiments on Ubuntu 20.04 with an AMD EPYC 7702 CPU and 512 GB of RAM. We built Chassis using Racket 8.12 and Rust 1.77.2. We also built Herbie with Racket 8.12 and Rust 1.77.2, using the most recent public release, Herbie 2.0.2.

Our evaluation uses all 547 benchmarks from the Herbie 2.0.2 release. The benchmarks are drawn from a variety of sources, including textbooks on scientific computing and numerical analysis, math libraries from various languages, and common mathematical kernels from geometry, hyperbolic geometry, statistics, and more. The benchmarks exhibit widely varying characteristics, ranging from those with significant accuracy problems to those with few accuracy problems, but less than desirable performance. This variation in source and characteristics allows us to demonstrate compilations along a wide spectrum of accuracy requirements and applications.

The 547 benchmarks are tested on 9 Chassis targets: three traditional ISAs (Arith, Arith+FMA, and AVX) three programming languages (C 99, Julia 1.10, and Python 3.10), and three software libraries (NumPy, CERN's vdt library, and Sun's fdlibm library).

## 6.1 Targets

| Target | Operators | L/E | S/V | Costs |
|---|---|---|---|---|
| Arith | + - * / $\sqrt{x}$ \|x\| | E | S | auto-tune |
| Arith+FMA | & fma | E | S | auto-tune |
| AVX | & fm*[10] rcp rsqrt | L | V | Fog [20] |
| C | math.h | L | S | auto-tune |
| Python | math[11] | E | S | auto-tune |
| Julia | Base[12] | E | S | auto-tune |
| NumPy | routines.math[13] | E | V | auto-tune |
| vdt | fast-*f, appr-*f [14] | L | S | auto-tune |
| fdlibm | math.h log1pmd | L | S | auto-tune |

**Figure 6.** Targets descriptions implemented for Chassis. For each, we list supported operators, whether the operators are modeled natively, whether it is scalar- or vector-style, and the source of the cost model.

Figure 6 lists the 9 target descriptions we developed and their characteristics, including the set of operators supported, whether those operators were linked (L) or emulated (E), whether conditionals used a scalar (S) or vector (V) cost model, and how the cost model was developed. The diversity of targets is clearly evident. All of the programming languages use a scalar-style evaluation, but both the AVX vector extensions for x86 and the NumPy library for Python use a vector style of evaluation where conditions are executed using masking (_mm256_blend_pd for AVX and numpy.where for NumPy). The targets that only provide accurate library functions (C, Python, Julia, NumPy) use emulated implementations for simplicity. The vdt and AVX libraries offer fast approximate operators; their targets link to the actual implementations, so that Chassis understands the accuracy implications of using these approximate operators. The C and AVX targets also offer both 32-bit and 64-bit operators, while other targets use 64-bit only. Almost all targets use auto-tuned cost models (see Section 7 for an evaluation of these cost models), but for the AVX target we used latency figures from Fog's well-known instruction tables [20].

The chosen targets exhibit widely varying characteristics. The Arith, Arith-FMA, and AVX targets don't offer any transcendental functions, while the other targets do. The Python and Julia targets have substantial interpretation overhead, meaning that operator costs are closely clustered, while other targets have starker divisions between fast and slow operations. Julia's target description includes a substantial number of high-accuracy functions such as cosd and log1p, while Python's target description does not even offer fma. The vdt library offers transcendental functions targeting 8 units in the last place of error, while other targets provide correctly-rounded or highly-accurate functions. The vdt library also offers a reciprocal square root function at two levels of accuracy, fast_isqrt and appr_isqrt. Meanwhile, the fdlibm library offers three variations on the logarithm function corresponding to subcomponents of its implementation. Simply put, Chassis makes it possible to describe a diverse set of targets with many differing features.

## 6.2 Comparison with Clang

Clang is a widely-used C compiler. We evaluate only Chassis' C target against Clang 14 on the 547 benchmarks from Herbie 2.0.2. Chassis' output is compiled using Clang with -O2 and the precise floating-point model; this is compared against six Clang optimization levels (-O0, -O1, -O2, -O3, -Os, -Oz), with and without fast-math, for a total of 12 configurations. For each compiled program, we measure its running time (on 10 000 pre-sampled input points) to compute the speedup relative to the original program compiled with -O0; we also compute the average accuracy measured with $p - \log_2$ ULP, where $p$ is the output precision of the expression. Note that, because the benchmarks have significant numerical accuracy issues, Clang's compiled output is fairly inaccurate. Also note that Clang produces a single output per configuration, while Chassis produces multiple outputs for each benchmark, covering a range of accuracy bounds. To aggregate over multiple benchmarks, we compute the geometric mean of speedups and the sum of accuracies to produce a joint Pareto curve.

Figure 7 contains the results. Chassis provides faster implementations across the entire spectrum of accuracy bounds, largely because it is able to combine accuracy-*improving* rewrites with much more aggressive performance-focused optimizations. Of course, Clang's semantics-preserving compilations preserves the exact source-level floating-point behavior, which Chassis does not. However, when benchmarks contain numerical inaccuracies, as in this case, semantics preservation merely means bug preservation. More interestingly, Clang's fast-math outputs are a closer comparison for Chassis. The fast-math configurations produce faster programs than standard configurations with little impact on accuracy, or even mild improvements. Despite this, Clang's optimizations are far more limited, so a developer willing to have the compiler change the source-level floating-point behavior is better served by Chassis than by Clang. For each benchmark, Chassis' run time averages a minute, while Clang takes less than a second.

---

[10] fmadd, fmsub, fmnadd, fmnsub
[11] https://docs.python.org/3.10/library/math.html
[12] https://docs.julialang.org/en/v1/base/math/
[13] https://numpy.org/doc/1.26/reference/routines.math.html
[14] exp, sin, cos, tan, tanh, log, log, asin, acos, atan

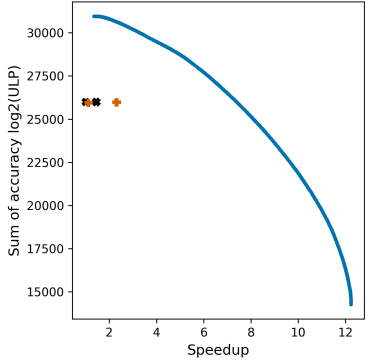

**Figure 7.** Chassis (blue circles) achieves better accuracy/performance to Clang both with (orange pluses) and without fast-math (black X's) at various optimization levels for all 547 benchmarks. We aggregate across the benchmarks by computing the geometric mean of speedups and summing accuracies to produce a joint Pareto curve. Up and right is better: the horizontal axis measures speedup relative to the input program compiled with -O0, and the vertical axis measures accuracy $p - \log_2 \text{ULP}$, where $p$ is the output precision of the expression.

## 6.3 Comparison with Herbie

Herbie is a widely-used numeric compiler [29]. Across all 9 targets and 547 benchmarks, we evaluate Chassis' target-specific output against Herbie's target-agnostic output, compiled to that target using the target-provided FPCore translation (see Section 4.2). Run time and accuracy are measured the same way as for Clang. We measure Chassis' speedup relative to the initial input programs, aggregating across benchmarks as before. On all targets, Chassis produces some outputs that are more accurate than Herbie's programs; to keep our comparison as conservative as possible and bias towards the baseline, we discard these outputs.

Figure 8 shows the results of this comparison. For all targets, Chassis produces target-specific programs that are faster, at a given accuracy bound, than the target-agnostic program produced by Herbie. In particular, Chassis achieves larger improvements for high-performance programs but less improvement for high-accuracy programs, which Herbie is especially tuned for. Note that the size of the gap differs based on the target. The hardware-focused targets (Arith, Arith+FMA, and AVX) show small gaps, on the order of several percentage points; for Arith+FMA and AVX, Chassis finds larger speedups for the most accurate programs. The small gaps are likely because the limited instruction sets of these targets mean fewer optimization opportunities exist, and also because many Herbie outputs use transcendental functions and are discarded. On the language-focused targets (C, Julia, and Python), Chassis achieves larger speedups. However, the speedups are notably smaller for Julia and

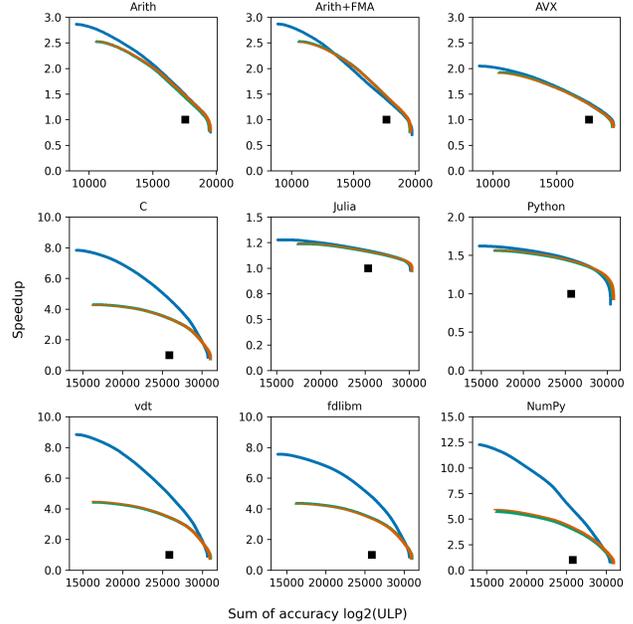

**Figure 8.** Chassis (blue circles) achieves better accuracy/speed tradeoffs than Herbie on its 9 targets for all 547 benchmarks. We aggregate across the benchmarks by computing the geometric mean of speedups and summing accuracies to produce a joint Pareto curve. Up and to the right is better: the vertical axis measures relative speedup of Chassis' programs over the input programs. For Herbie's programs, we desugar unsupported operators (orange dashes) or transcribe directly (green "tri"s); unsupported programs are ignored. The black square represents the (aggregate) accuracy and speed of the input programs (Speedup = 1).

Python. We hypothesize that this is for two reasons: the large overheads of these languages lead to flat cost models, so Chassis' benefits less from cost information; and also, these targets don't provide low-accuracy accelerators or multiple precisions that Chassis can use to improve performance. Julia also provides high-accuracy helper functions that Chassis *is* able to leverage to provide higher-accuracy options than Herbie. While we discard those outputs in our evaluation, they are still valuable for users. Finally, the library-focused targets (NumPy, fdlibm, vdt) show dramatic speedups, up to 1.9× for vdt. This shows the value of choosing approximate library functions, like the fast_ and appr_ functions offered by vdt, when accuracy permits. Chassis and Herbie have similar run times.

Observe that Herbie can find faster programs than Chassis' for a given accuracy bound, especially for high accuracy programs. Figure 9 presents an alternative view of Figure 8: the vertical axis now represents the (aggregate) speedup over Herbie's output programs for a given accuracy rather than over the initial input programs. Note that this plot, unlike Figure 8, is highly sensitive to outliers. Chassis fails to find similar programs to the highest-accuracy programs that Herbie finds in a handful of benchmarks (19 of 547, about 3.5% of the total); the result is the set of "tails" on the right-hand side of most of the plots in Figure 9. We suspect a number of possible causes. Chassis' equality saturation instances are more intensive due to mixed real-float expressions and are likely more impacted by resource limits. In addition, Chassis is more sensitive to its cost models than Herbie, as Chassis relies on its cost models for heuristics (Section 5.2), e-graph extraction (Section 5.1), and more.

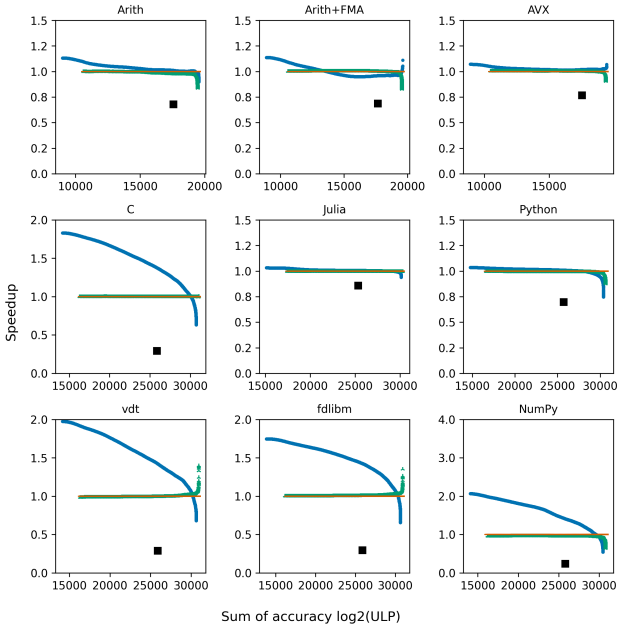

**Figure 9.** Chassis achieves better accuracy/speed tradeoffs than Herbie on its 9 targets for all 547 benchmarks. This figure shows the same data as Figure 8, with the vertical axis measuring speedup over Herbie's output programs for a given accuracy rather than the input programs. Up is better.

It is not always clear how to compare target-agnostic and target-specific programs; we attempt to always bias our results toward Herbie. When a Herbie output uses an operation that is unavailable on a given target, we attempt to desugar to simpler functions; for example, on the Python target, we replace $\text{fma}(x, y, z)$ with $x * y + z$. Chassis produces *target-specific* programs, so it never uses unsupported operators and thus doesn't need this step. The impact of this desugaring can be seen in the gap between orange (desugaring) and green (no desugaring) dots in Figure 8, which is small compared to the gap between Chassis and Herbie. If an operator is still unsupported after desugaring, typically because an operation is fundamentally missing, we discard this Herbie output from consideration and construct the Pareto curve from Herbie's other outputs. If all Herbie outputs are discarded for a given benchmark, we remove it from consideration both for Herbie and for Chassis. This significantly biases the results toward Herbie because these are typically the most challenging benchmarks to implement on a particular target.

### 6.4 Case Studies

We analyze three representative case studies drawn from the 547 benchmarks from Herbie 2.0.2. They demonstrate how Chassis uses target-specific features to select fast but inaccurate operators, accurate but slow operators, or even new accelerators, which a developer may be interested in making visible. Moreover, Chassis finds these target-specific rewritings *solely* from information provided by the target descriptions. For each case study, we compare Chassis' target-specific programs to Herbie's uncustomized programs.

**Quadratic Formula.** The first case study is a modification of the usual quadratic formula.

$$\left(-b_2 + \sqrt{b_2^2 - ac}\right)/a$$

Compared to its usual form $\left(-b + \sqrt{b^2 - 4ac}\right)/2a$, the expression above uses as an argument $b_2 = b/2$. We omit branches where Chassis primarily applies polynomial approximation to implement the expression.

Herbie produces an implementation with two branches of interest:

$$\begin{cases}(\text{sqrt}(b_2 * b_2 - a * c) - b_2)/a, & \text{if } -10^{132} < b_2 \leq 0, \\ c/(-b_2 - \text{sqrt}(-(a * c - b_2 * b_2))), & \text{if } 0 < b_2 \leq 10^{127}.\end{cases}$$

On the other hand, Chassis, when targeting AVX, generates an implementation with similar branches, but leverages the many fma variants available. It implements the first branch with

$$(\text{sqrt}(\text{fma}(b_2, b_2, \text{fnmadd}(a, c, 0))) - b_2)/a.$$

and the second branch as

$$\frac{c * a}{a * \text{fmsub}(b_2, -1, \text{sqrt}(\text{fnmadd}(a, c, \text{fma}(b_2, b_2, 0))))}.$$

(Notice that Chassis failed to simplify extra multiplications by $a$ on both numerator and denominator. Missing simplifications often indicate that resource limits were hit during equality saturation before the simplification could be found.) Importantly, in single-precision, Chassis can also use a fast approximate reciprocal instruction rcpss, so the second branch becomes

$c * \text{rcpss}(\text{fmsub}(b_2, -1, \text{sqrt}(\text{fnmadd}(a, c, \text{fma}(b_2, b_2, 0)))))$.

(In contrast to the double-precision expression, Chassis avoids redundant multiplications by $a$.)

**Ellipse Angle.** This case study considers the following formula.
$$a^2 \sin^2(\pi/180 \cdot \theta_{\text{deg}}) + b^2 \cos^2(\pi/180 \cdot \theta_{\text{deg}})$$
The expression computes the coefficient $A$ of an ellipse's implicit equation $Ax^2 + Bxy + Cy^2 + Dx + Ey + F = 0$ (assuming $B^2 - 4AC < 0$) from the rotation angle $\theta$ and axis lengths $a$ and $b$. For this expression, the argument is in degrees rather than radians.

Herbie struggles to improve this expression, choosing various series expansions of sin and cos with little success. Chassis does better when targeting Julia. In particular, Chassis produces a more accurate implementation than the one Herbie produces by using Julia's trigonometric functions in degrees:
$$(a * \texttt{sind}(\theta_{\text{deg}}))^2 + (b * \texttt{cosd}(\theta_{\text{deg}}))^2.$$
Chassis also produces a slightly less accurate implementation with an odd branch of the form
$$(\texttt{deg2rad}(\theta_{\text{deg}}) * a)^2 + (b * \texttt{abs2}(\texttt{cosd}(\theta_{\text{deg}}))) * b$$
where deg2rad is the usual degree-to-radian transformation, and abs2 computes $|x|^2$, the absolute square of a number.

**Inverse Hyperbolic Cotangent.** Recall the expression from the overview example.
$$\coth^{-1}(x) = \frac{1}{2} \cdot \log\left(\frac{1+x}{1-x}\right)$$
The expression computes the inverse hyperbolic cotangent from the hyperbolic angle measure $x$. For small $x$, the expression suffers from numerical error.

Herbie discovers a highly accurate implementation by using the special numeric function log1p:
$$0.5 * (\texttt{log1p}(x) - \texttt{log1p}(-x)).$$
On the other hand, Chassis, optimizing for the fdlibm target, leverages the special log1pmd operator, implementing the expression as
$$\texttt{log1pmd}(x) * 0.5.$$
Thus, Chassis shows that making the operator visible could improve the accuracy and performance of other numerical programs.

## 7 Limitations and Future Work

While Chassis improves substantially over either numeric or traditional compilers, it could likely do better with more accurate cost models. Figure 10 shows the correlation between estimated and actual run time over the 547 benchmarks of Section 6. The correlation is positive, but a number of exceptional points are clearly visible. These stem from a variety of issues that could be resolved with a better cost model. For example, many operators have input-dependent costs, with hardware instructions often slowing down significantly in the presence of denormal numbers, or languages like Python throwing exceptions for divisions by zero. Software library functions that use lookup tables can also show cache effects that depend on the application at large. We are interested in improving Chassis' cost models, though there are significant constraints due to their use during e-graph extraction.

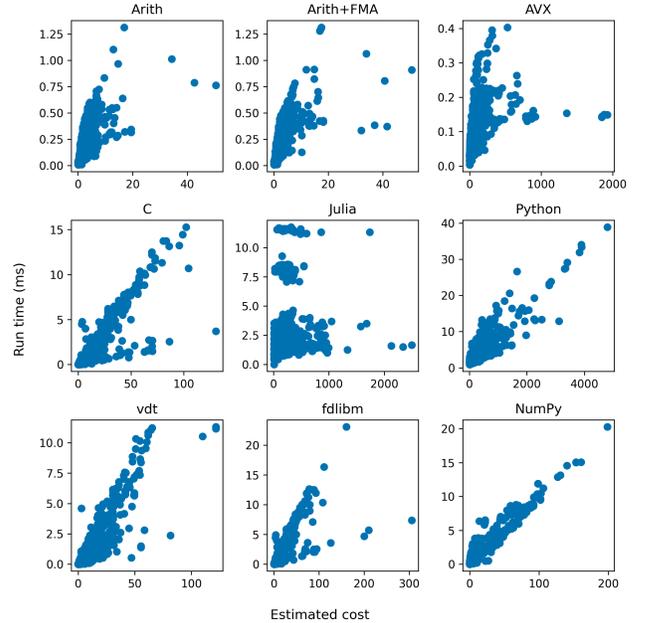

**Figure 10.** Chassis' cost model reasonably approximates the actual run time of candidate implementations. Across all 547 benchmarks and 9 targets, we compare Chassis' estimated cost versus run time of each output floating-point program. We observe a moderate-to-strong correlation.

It would also be interesting to consider accuracy as well as cost during extraction, potentially removing the need to run an iterative improvement loop. The challenge is that accuracy is difficult to evaluate one e-node at a time, as necessary for e-graph extraction. We plan to explore evaluating Chassis' local error heuristic over e-graphs and using that to guide extraction. Beyond numerical compilers, we hope that our novel typed extraction algorithm finds applications in other equality saturation problems.

## 8 Related Work

Chassis is most similar to Herbie [29], a repair tool which combines rewrite rules, series expansions, and a regime inference algorithm to improve numerical accuracy. The follow-on Pareto-Herbie tool [31] adds precision tuning and multiple Pareto-optimal outputs. Chassis outperforms Herbie, finding better speed/accuracy tradeoffs. Salsa [10] is a similar tool that improves numerical accuracy through rewrites, but uses abstract interpretation to safely approximate round-off errors and guide its rewrites. Daisy [14] uses genetic programming, again combining term rewriting and precision tuning. Other tools use precision tuning alone to meet accuracy bounds: FPTuner [8] combines symbolic Taylor expansions with quadratic programming, Precimonious [30] uses delta debugging, and HiFPTuner [22] uses hierarchical search. Genetic programming and SMT-based algorithms have also been applied to fixed-point program synthesis for polynomials [13, 18]. Meanwhile, the STOKE [32, 33] superoptimizer stochastically rewrites programs via Monte Carlo sampling to find faster programs under some accuracy bound.

Prior work has also explored a compilation perspective. Rosa [12] takes real-valued programs, input preconditions, and an accuracy bound, and synthesizes code that meets that bound, largely through precision tuning. It verifies that the error bound is met using an SMT solver. Spiral [21] is a specialized compiler for mapping linear algebra kernels to highly efficient, target-specific C, Fortran, or Verilog implementations using a rewrite system tuned to handle Fourier transforms, tensor algebra, image processing, and more. In verified compilation, the Icing language [2] features a compiler that formalizes fast-math optimizations and applies them only when it is provably safe to do so. For Systems-on-Chip (SoC) design, prior work [23] proposes Instruction Level Abstraction (ILA), in which accelerators are modeled as real operators to better reason about their interactions with programs.

Detecting numerical error is also a well-studied topic. Herbie [29] introduces local error, an approximate metric of an operator's individual contribution to numerical error, and uses it to guide rewrites. HiFPTuner [22] uses community detection and dependence analysis to identify operators and variables where precision-tuning could be applied. FPDebug [3] and PositDebug [9] perform shadow execution along with high-precision computation to detect numerical error, the former using Valgrind and the latter with compile-time instrumentation. Using a slightly different approach, Shaman [16] relies on error-free transformations (EFTs) to dynamically track numerical error by replacing floating-point numbers with more accurate values from the EFTs, recording the number of significant digits in a computation. FPTaylor [34] uses a technique called symbolic Taylor expansions to compute tight error bounds of floating-point expressions, while Satire [15] improves on this technique by applying path strength reduction and bound optimization. Meanwhile, Real2Float [26] applies semi-definite programming to bound round-off errors of polynomials, PRECiSA [36] calculates formally-verified symbolic error bounds for floating-point expressions, and Rosa [12] uses an SMT-based approach to soundly estimate error bounds.

Finally, automated selection and synthesis of floating-point operator implementations is a new yet highly active area of research. OpTuner [5] optimally selects the best implementations of mathematical functions at each call site using error Taylor series and integer linear programming. MetaLibm [25] provides parameterized operator implementations, and MegaLibm [6] provides a framework for designing and synthesizing math library implementations from scratch. RLibM [1] synthesizes correctly-rounded math library functions using counterexample-guided polynomial generation and linear programming. Chassis does not synthesize implementations, but proposing new operators for a target would be an interesting future step.

## 9 Conclusion

This paper combines accuracy-aware rewriting techniques from numerical compilers with target-aware instruction selection from traditional compilers to create Chassis, a target-aware numerical compiler. Chassis contributes a target description language that relates floating-point instructions to the mathematical formula they approximate. This correspondence permits a novel algorithm for instruction selection modulo equivalence, which applies equality saturation to select target-specific instructions. To guide this heavyweight optimization pass, Chassis uses new accuracy- and cost-aware heuristics. We also introduce a new typed-extraction algorithm for navigating mixed-type equality saturation problems. We evaluate Chassis on a diverse set of 9 targets, achieving better accuracy and performance than state-of-the-art traditional and numerical compilers by leveraging target-specific characteristics. We hope this work inspires future research that melds traditional compiler techniques with automated numerical analysis.

## Acknowledgements

We would like to thank our anonymous reviewers for their constructive feedback towards preparing the final version of this paper. We especially acknowledge Guy Steele, for his thoughtful guidance, wise insights, and valuable time throughout the shepherding process. We thank John Regehr for providing excellent feedback on an early draft of this paper. We thank Bill Zorn for his deep knowledge of computer number systems and applications of Chassis in industry. We also thank Chandrakana Nandi, Edward Misback, Amy Zhu, and Thia Richey for their time reviewing our paper. Finally, we thank the entire Herbie team for contributing code to the Chassis project and their continual support.

This material is based upon work supported by the U.S. Department of Energy, Office of Science, Office of Advanced Scientific Computing Research, ComPort: Rigorous Testing Methods to Safeguard Software Porting, under Award Number DE-SC0022081. This work was also supported by the National Science Foundation under Grant Nos. 2346395, 2232339. Any opinions, findings, and conclusions or recommendations expressed in this material are those of the authors and do not necessarily reflect the views of the National Science Foundation.